\documentclass{article}

\usepackage{arxiv}

\usepackage[utf8]{inputenc} % allow utf-8 input
\usepackage[T1]{fontenc}    % use 8-bit T1 fonts
\usepackage{hyperref}       % hyperlinks
\usepackage{url}            % simple URL typesetting
\usepackage{booktabs}       % professional-quality tables
\usepackage{amsfonts}       % blackboard math symbols
\usepackage{nicefrac}       % compact symbols for 1/2, etc.
\usepackage{microtype}      % microtypography
\usepackage{lipsum}		% Can be removed after putting your text content
\usepackage{graphicx}
\usepackage{natbib}
\usepackage{doi}
\usepackage{textcomp}
\usepackage{times}
\usepackage{setspace}
\usepackage{makecell}
\usepackage[font=small,compatibility=false]{caption}
\usepackage{listings}

\usepackage{amsmath}% http://ctan.org/pkg/amsmath

\usepackage[labelfont=bf]{caption}

\title{Socio-Political Feedback on the Path to Net Zero}

\author{ \href{https://orcid.org/0000-0001-6382-1381}{\includegraphics[scale=0.06]{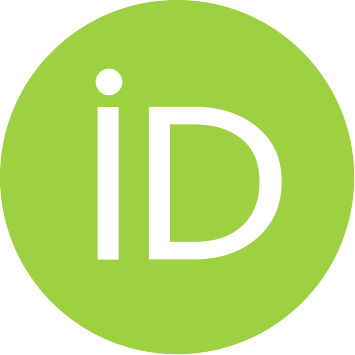}\hspace{1mm}Saverio Perri} \\
	High Meadows Environmental Institute,\\
	Princeton University,\\
	Guyot Hall, Princeton,\\
	NJ 08544, USA.
	\texttt{sperri@princeton.edu} \\
	\And
	\href{https://orcid.org/0000-0002-8216-5639}{\includegraphics[scale=0.06]{orcid.pdf}\hspace{1mm}Simon Levin} \\
	Ecology \& Evolutionary Biology, \\
	Princeton University,\\
	106A Guyot Ln, Princeton, \\
	NJ 08544, USA.
	\texttt{slevin@princeton.edu} \\
	\And
	\href{https://orcid.org/0000-0001-7992-1774}{\includegraphics[scale=0.06]{orcid.pdf}\hspace{1mm}Lars O. Hedin} \\
	Ecology \& Evolutionary Biology, \\
	Princeton University,\\
	106A Guyot Ln, Princeton, \\
	NJ 08544, USA.
	\texttt{lhedin@princeton.edu} \\
	\And
	\href{https://orcid.org/0000-0002-3566-323X}{\includegraphics[scale=0.06]{orcid.pdf}\hspace{1mm}Nico Wunderling} \\
	Earth System Analysis,\\
	Potsdam Institute for Climate Impact Research (PIK),\\
	Potsdam, Germany.
	\texttt{nico.wunderling@pik-potsdam.de} \\
	\And
	\href{https://orcid.org/0000-0001-9378-207X}{\includegraphics[scale=0.06]{orcid.pdf}\hspace{1mm}Amilcare Porporato} \\
	Department of Civil and Environmental Engineering, \\
	Princeton University,\\
	Princeton, NJ, USA.
	\texttt{aporpora@princeton.edu} \\
}

\renewcommand{\shorttitle}{\textit{arXiv} Template}

\hypersetup{
pdftitle={Socio-Political Feedback on the Path to Net Zero},
pdfsubject={q-bio.NC, q-bio.QM},
pdfauthor={Saverio Perri, Simon Levin, Lars O. Hedin, Nico Wunderling, Amilcare Porporato},
pdfkeywords={Net zero, Socio-political acceptance, Anthropogenic emissions, Clean energy, International coordination, Prisoner's dilemma},
}

\begin{document}
\maketitle

\begin{abstract}
Anthropogenic emissions of CO$_2$ must soon approach net-zero to stabilize the global mean temperature. Although several international agreements have advocated for coordinated climate actions, their implementation has remained below expectations. One of the main challenges of international cooperation is the different degrees of socio-political acceptance of decarbonization.
Here we interrogate a minimalistic model of the coupled natural-human system representing the impact of such socio-political acceptance on clean energy investments and the path to net-zero emissions.
We show that incentives and carbon pricing are essential tools to achieve net-zero before critical CO$_2$ concentrations are reached, and deep international coordination is necessary for a rapid and effective transition. Although a perfect coordination scenario remains unlikely, as investments in clean energy are ultimately limited by myopic economic strategies and a policy system that promotes free-riding, more realistic decentralized cooperation with partial efforts from each actor could still lead to significant emissions cuts.
\end{abstract}

% keywords can be removed

\section*{Introduction}
Greenhouse gas emissions from fossil-fuel combustion are warming the Earth's climate at an unprecedented rate \citep{field2004global,canadell2007contributions}, potentially leading to the greatest tragedy of the commons in human history \citep{gardiner2011perfect,battersby2017news}. To limit the risk of crossing critical climate thresholds and achieve the goals set by international agreements \citep{rockstrom2009planetary,agreement2015paris,steffen2015planetary,schellnhuber2016right,bodansky2017international,steffen2018trajectories}, global carbon emissions should reach net zero (NZ) by 2050. This requires facing not only technological issues but also social, economic, and political challenges related to the hindrance to change in  economies that depend on fossil fuel \citep{levin2012overcoming,davis2018net,preiser2018social,garmestani2019untapped,lilliestam2021effect,Moore2022Determinants}. 
Across countries and pluralistic societies, a combination of diverging economic priorities, diverse perceived impacts of climate change, and a broad spectrum of sensitivities to environmental issues makes rising carbon emissions a wicked problem, causing disagreement about its solution \citep{rittel1973dilemmas,incropera2016climate}.

On a fundamental level, the dynamics and feedbacks between socio-political acceptance, global investment in clean energy, and, consequently, the achievement of NZ emissions remain unclear. In this paper, we investigate the linkages between actions and reactions in the coupled socio-political-climate system. We explore the role of such feedbacks on the perturbation of long-term trends in CO$_2$ concentration caused by the global energy demand and the use of either fossil fuels or clean energy to meet it.

While financial incentives and carbon pricing are often presented as the main strategies to facilitate emissions reduction \citep{cox2016financial,baranzini2017carbon,stiglitz2017report}, global climate policies have neglected {\it de facto} the highly contextual and socio-political nature of decarbonization \citep{keohane2016cooperation,rosenbloom2020opinion}.
International climate agreements leverage emissions cuts as a long-term global public good \citep{kaul1999global}, but deep coordination has been hindered by priorities that are expressed at the scale of local authorities and that include socio-political influence in addition to tangible short-term economic factors \citep{nordhaus2020climate}.
The connection between actions and gain becomes tenuous for the climate problem, where the threat, when perceived, is often seen as a long-term global issue more than an impending local danger. As a result, the propensity of a country to collaborate with others to reduce emissions depends on the socio-political perception of climate actions.

Countries have delayed taking real action for decades, quarreling over costs and responsibility \citep{lamb2020discourses}. In such a situation, each actor is enticed to take advantage of others' efforts without directly contributing to their action (free-riders) \citep{keohane2016cooperation,nordhaus2020climate}. This is particularly true in developing and conservative countries, where emissions reduction can be perceived as disrupting the national economy \citep{holtz1995stoking,nordhaus2001global}. This `prisoner's dilemma' faced by  climate actors \citep{rapoport1965prisoner,soroos1994global} promises higher rewards for betraying, thus promoting a free-rider behavior rather than cooperation by investing in clean energy. Yet, if no one invests in clean energy, the penalty is far worse than the share to pay for cooperating \citep{putnam1988diplomacy}. 

The fear of materializing this dangerous penalty may turn the prisoners' dilemma into a coordination game \citep{barrett2012climate,decanio2013game}, where the best outcome for everyone unequivocally comes from cooperation. But the different perceptions of climate change risk, which are also related to significant uncertainties in identifying critical thresholds in the climate system, can easily promote skepticism and inhibit coordination \citep{barrett2012climate}. Whether international climate negotiations develop under a prisoners' dilemma or coordination game scheme may thus evolve depending on time-varying socio-political processes interconnected to strategies on emission cuts.
%%%%%%%%%%%%%%%%%%%%%%%%%%%%%%%%%%%%%%%%%%%%%%%%%%%%%%%
%%%%%
%%%%%%%%%%%%%%%%%%%%%%%%%%%%%%%%%%%%%%%%%%%%%%%%%%%%%%%
\vspace{-5pt}
\section*{Results and Discussion}
\label{Results}
%================================================================%
% Socio-political feedback on decarbonization
%================================================================%
\subsection*{Socio-political feedback on decarbonization}
\label{sec:Dynamics}
Although the impact of socio-political acceptance on clean energy transitions and other actions to address climate change has been widely analyzed from both conceptual and economic points of view \citep{devine2005beyond,wolsink2020distributed,constantino2021cognition,willner2021investment}, their effect are treated as exogenous in climate studies \citep{beckage2020earth,Moore2022Determinants}. 
Here, we propose a parsimonious representation of how energy demand perturbs the global carbon cycle while explicitly accounting for the interdependence between the perceived impact of climate change and mitigation actions (see Methods).

\begin{figure} [ht!]
   \centering 
\noindent\includegraphics[width=24pc]{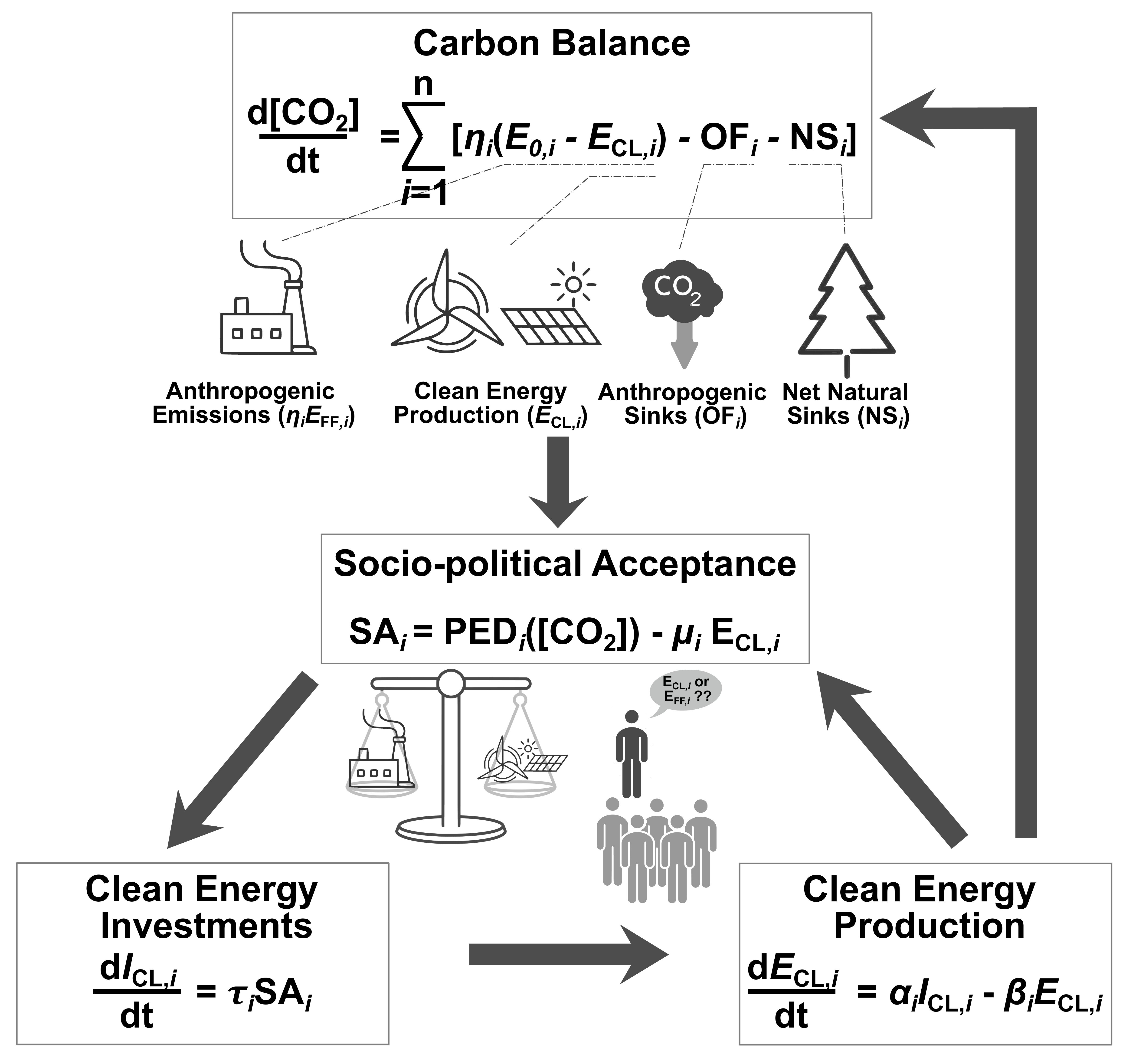}
   \caption{\textbf{Socio-political Feedback on Net Zero Emissions.} Schematic representation of the perturbation of the global carbon cycle due to energy demand and fossil fuels combustion and its feedback on clean energy investments as mediated by the socio-political system. For each block $i$, the socio-political acceptance (SA$_i$) exerts a feedback on the carbon balance perturbation through clean energy investments, which, in turn, affects SA$_i$ directly (increasing clean energy production) and indirectly (reducing CO$_2$ emissions; see Methods for details).}
  \label{fig:Fig_1}
\end{figure}

Our framework accounts for the essential feedback of socio-political acceptance and incentives to promote decarbonization on the path to NZ. We link country decisions to invest in clean energy on a) socio-political acceptance that is a function of the perceived impact of climate change and investment in clean energy, and b) the potential to incentivize the transition through financial measures and carbon pricing. To explore these issues quantitatively, we set up a dynamical model (see Methods and Figure \ref{fig:Fig_1}) aiming at describing fluctuations in atmospheric CO$_2$ driven by the global energy demand, which either fossil fuels or clean energy could meet. Finally, different scenarios of collaborations among countries are analyzed to emphasize the importance of a strong coordinated effort to achieve NZ rapidly and before critical CO$_2$ concentrations are crossed.

The increase in atmospheric CO$_2$ due to the combustion of fossil fuels can be mitigated by increasing the share of clean energy used to meet the global energy demand or partially compensated by offset technologies (e.g., carbon capture and storage, direct air capture, nature-based solutions) \citep{ciais2014carbon,friedlingstein2021global}. The decision of whether investing in clean energy to reduce emissions is assumed to be a function of incentives to promote decarbonization or carbon pricing and socio-political acceptance.
The latter depends on the damage potentially caused by rising CO$_2$ and the perceived impact of increasing low-emission energy production. 

Whereas increasing CO$_2$ concentration is associated with global warming, air pollution, and more frequent heatwaves, droughts, and floods \citep{van2006impacts,schiermeier2018droughts}, its consequences are far from immediate and extremely variable in space.
For example, it has been estimated that the time lag between a CO$_2$ emission pulse and its maximum impact on average temperature may be of several decades \citep{ricke2014maximum,zickfeld2015time,allen2018solution}. Moreover, the vulnerability of social and ecological communities displays high spatial variability with certain regions much more threatened by climate change than others \citep{reid2009mapping,flanagan2011social,field2014climate,spangler2019characterizing}.  
Such high spatio-temporal variability and uncertainty contribute not only to a diverse resilience to climate change but also to different public risk perceptions \citep{leiserowitz2006climate}, and climate negotiation strategies may adjust accordingly\ cite{barrett2012climate}. In spite of this convolution, when CO$_2$ concentration grows, so does the perceived socio-economical damage, and it becomes more and more acceptable to invest in clean energy to achieve NZ emissions \citep{ungar1992rise,nath2011critical,ricke2014natural,Moore2022Determinants}. 
However, this NZ goal is often perceived as a challenge to employment, especially in fossil fuel-based economies, and its acceptance is expected to decrease with booming clean energy \citep{longo2008internalization,lambert2012challenges}. For given socio-political acceptance, governments can decide to boost the investments through financial incentives and by introducing carbon pricing, which directly and indirectly promote sustainable energy sources \citep{campiglio2016beyond,baranzini2017carbon}. 

\begin{figure} [ht!]
   \centering 
\noindent\includegraphics[width=.75\linewidth]{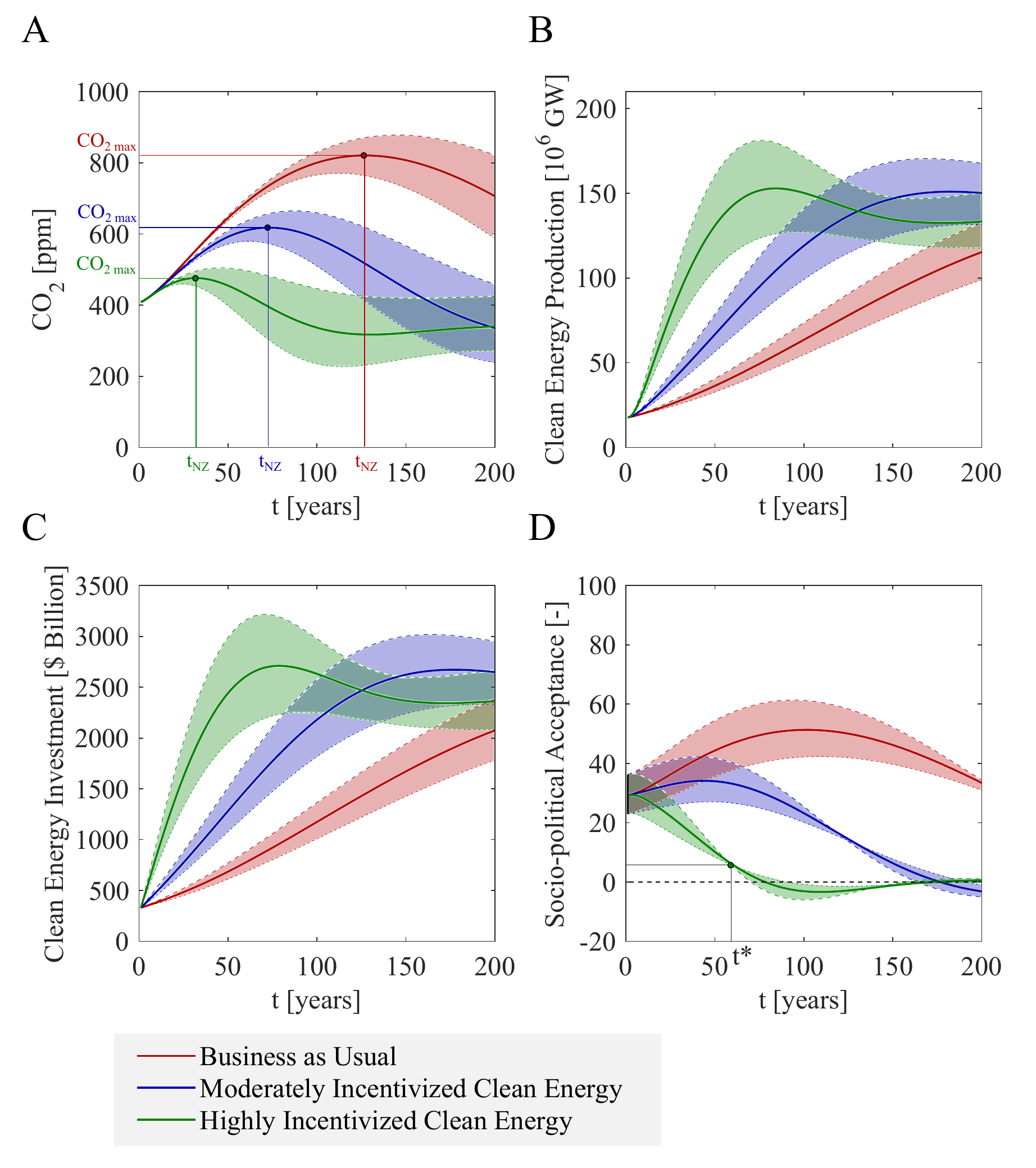}
% \end{figure}
% \clearpage
{\begin{center}
\captionof{figure}{{\textbf{Clean energy transition.} Simulated global (A) atmospheric CO$_2$ concentration, (B) clean energy production, (C) clean energy investment, (D) socio-political acceptability under different scenarios of carbon pricing and financial incentives and in case of deep coordination. The green curves refer to highly incentivized clean energy, while the blue and red are obtained for moderately incentivized and disincentivized clean energy, respectively. The shaded areas show the sensitivity of the decarbonization trajectories to the global perception of the impact of rising [CO$_2$] and $E_{CL}$. The upper limits (dashed lines) are obtained for low perceived risk of increasing carbon dioxide in the atmosphere and elevated perceived negative impact of clean energy. The lower limit (dashed-dotted lines) depicts a situation in which the socio-political system is keen to invest in clean energy due to the common perception that its benefits outweigh its drawbacks. The time t$^*$ represents a hypothetical time scale over which the socio-political acceptance of a conservative SPS becomes higher than that of a green SPS. The model parameters used in these simulation can be found in the Extended Data Table 1.}}
  \label{fig:Fig_2}
\end{center}}
%\textbf{The parameters used for these simulations are...}
%   \label{fig:Fig_2}
  \end{figure}

Our modeling framework links the concept of socio-political acceptance of decarbonization to energy-driven carbon emissions and consequent carbon cycle perturbation (see Methods). Our results (Figure \ref{fig:Fig_2}) show that incentives (or penalties) can have a significant impact on global CO$_2$ dynamics (Figure \ref{fig:Fig_2}A) by driving clean energy production (Figure \ref{fig:Fig_2}B) and investments (Figure \ref{fig:Fig_2}C). Highly incentivized decarbonization (green curves) leads to a more rapid achievement of NZ. As stakeholders reduce the incentives to encourage clean energy investments (blue and red curves), it takes more time to reach NZ. Moreover, NZ is achieved at higher levels of CO$_2$, which could be too elevated to respect international agreements and reduce the risk of crossing climate tipping points \citep{rockstrom2009planetary}.

%%%%%%%%%%%%%%%%% Begin Figure 2 %%%%%%%%%%%%%%%%%%%%
%   \vspace{1cm}
%%%%%%%%%%%%%%%%% End Figure 2 %%%%%%%%%%%%%%%%%%%%
Figure \ref{fig:Fig_2} also demonstrates that policies can significantly boost clean energy investments, but these are ultimately limited by the public perception of their trade-off between costs and benefits. The shaded areas in Figure \ref{fig:Fig_2} show the sensitivity of the decarbonization trajectories to the global perception of the impact of rising CO$_2$ and clean energy. This perception defines what we call here hindrance to change ($\mu$ parameter in the model; see Methods). The upper limit of such dynamics (higher concentrations represented by dashed lines in Figure \ref{fig:Fig_2}A) is dictated by the low perceived risk of increasing carbon dioxide in the atmosphere and elevated perceived negative impact of clean energy. This high hindrance scenario is here described, without reference to specific political movements, as conservative socio-political system (SPS), as it tends to preserve the \textit{status quo} of a fossil-fuel-dependent economy rather than supporting a transition to clean energy solutions. 
The lower trajectory (dashed-dotted lines) depicts a situation in which the SPS is keen to invest in clean energy due to the common perception that its benefits outweigh its drawbacks. This low hindrance scenario will be referred to as green SPS.

For a given SPS, the socio-political acceptance dynamically responds to fluctuations in CO$_2$ and clean energy (Figure \ref{fig:Fig_2}D). Increasing   CO$_2$ emissions lead to temperature increments and human health decline due to air pollutants and associated aerosols \citep{emanuel2018we,markandya2018health,scovronick2019impact}. As a result, when CO$_2$ rises, so does the propensity of an SPS to invest in decarbonization. However, as clean energy starts meeting a significant portion of the global demand and atmospheric CO$_2$ concentration declines, further investing in decarbonization becomes less and less acceptable. This negative feedback may be reinforced by the perceived negative implications of moving away from a fossil-fuel-dependent economy.

If the socio-political acceptance keeps declining and becomes negative, the investment in clean energy drops, and emissions may re-intensify. Increasing CO$_2$ would, however, raise again public concern about climate alterations that would result in new investments in decarbonization. This type of dynamics indicates the possibility of an `overshooting' in the investments (see green curves in \ref{fig:Fig_2}A), meaning that strong incentives can bring clean energy production above a socially acceptable level and the CO$_2$ below the equilibrium point (i.e., steady-state). From that point, the reduction in incentives would lead to subsiding investments until a stable demand/supply is attained.

When the global energy demand stabilizes, the acceptance eventually approaches zero, clean energy remains stable and CO$_2$ reaches steady-state.
However, reaching NZ and stabilizing CO$_2$ may not be sufficient to mitigate climate change. If a significant portion of the total energy demand will still be satisfied by fossil fuels, the steady-state CO$_2$ may remain too high to prevent global warming. Lower concentrations could be achieved by reducing the total energy demand (Extended Data Figure 1A) and increasing carbon capture and sequestration (Extended Data Figure 1B). Moreover, a more efficient energy conversion of fossil fuels can significantly reduce emissions. Interestingly, the steady-state CO$_2$ drops significantly if the SPS’s sensitivity to increasing emissions is high (see Methods). 

Remarkably, the relative weight of positive and negative feedbacks between climate action and the socio-political acceptance can change over time depending on CO$_2$ concentrations and clean energy production. Such perception shift occurs if early investments of the green SPS cause a rapid decrease in CO$_2$ and a rushed conversion to low-carbon energy solutions, which ultimately reduces the acceptance of further investments (Figure \ref{fig:Fig_2}D). The conservative SPS, on the contrary, tends to delay the deployment of clean energy and, as a consequence, the reduction in CO$_2$ emissions is slower than in the green SPS. If the difference in these trends is large enough, there could be a point in time (t$^*$ in Figure \ref{fig:Fig_2}D) in which the acceptance is virtually higher in the conservative SPS than in the green SPS. However, it is important to stress that this shift only emerges from comparing different trajectories under scenarios of deep global collaboration. As better explained in the following sections, the coexistence of conservative and green SPS leads to different levels of investment in response to the global CO$_2$ and the local impact of clean energy.
These findings confirm that one of the major challenges of implementing large-scale decarbonization is influencing social norms and behavior to make it politically acceptable \citep{weber2015climate,constantino2021source}.

It is also worth noting that maintaining CO$_2$ concentrations below critical levels requires investing in clean energy even when its acceptability is low. This is particularly challenging in developing countries that have limited resources and where the public opinion is less sensitive to environmental issues than developed countries \citep{mertz2009adaptation}. Pushing the investments below the acceptable threshold (i.e., negative socio-political acceptance in Figure \ref{fig:Fig_2}D) can result in a gradual reduction of the investments.
The country-specific or local disparities in socio-political acceptance of clean energy transition raise the question of whether an international collaboration where all entities put equal effort is even possible. 
%======================%
%      Results
%======================%
\section*{Coordination, cooperation, or defection}%Achieving NZ under different coordination, cooperation, and defection scenarios
\label{sec:Games}
Effective climate change mitigation requires deep international coordination \citep{keohane2016cooperation}. Ideally, climate negotiations should follow the pattern of a coordination game where decision-makers enforce cooperative behavior \citep{branzei2008models,barrett2012climate}. A similar, but perhaps more realistic scheme, is the one of a bargaining game where players negotiate compatible demands \citep{smead2014bargaining}.
Until now, however, in spite of the ambitious goals set by the Paris agreement and previous international treaties \citep{quarrie1992earth,protocol1997kyoto,agreement2015paris}, deep coordination has rarely been observed \citep{victor2011global}. As each player feels enticed to defect against the others to gain a short-term economic advantage \citep{krasner1999sovereignty,keohane2016cooperation}, international agreements have mostly generated prisoners' dilemmas \citep{magli2022coordination}. 

The typical pattern witnessed so far depicts a wide range of commitments across nations and local entities with failure to promote self-enforcing international collaboration \citep{barrett1994self,nordhaus2020climate}. However, as pointed out by Keohane and Victor 2016 \citep{keohane2016cooperation}, even this shallow coordination, defined as cooperation, could lead to substantial cuts in emissions. The question is then, how effective is cooperation compared to coordination in achieving NZ? Also, what if a substantial number of countries decide to act for short-term national advantages and ignore international agreements on emissions cuts in what is known as the defection scenario? 
%%
%%%%%%%%%%%%%% Begin Figure 3 %%%%%%%%%%%%%%%%%%%%%
\begin{figure} [h]
   \centering 
\noindent\includegraphics[width=.45\linewidth]{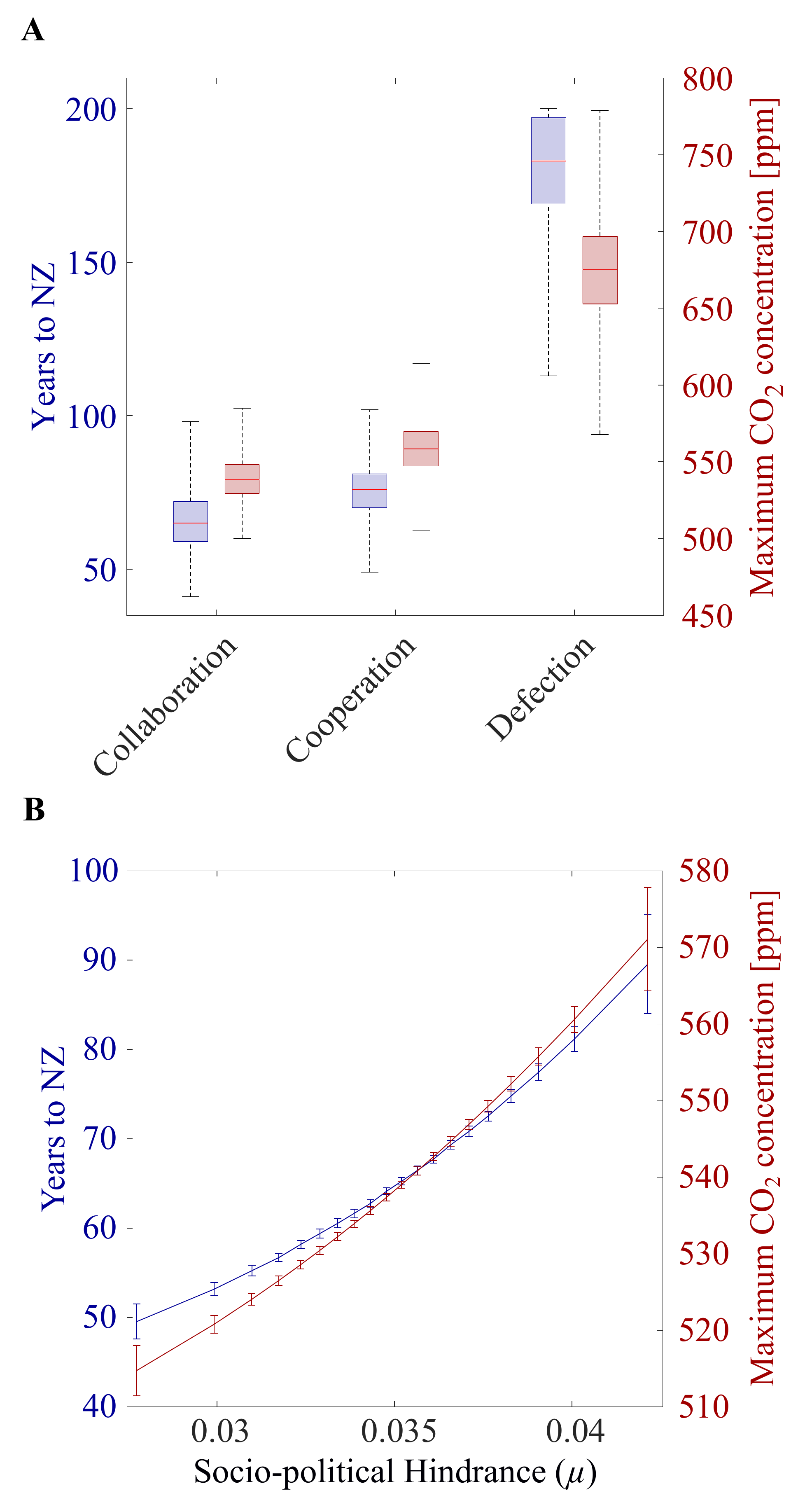}
  \caption{\textbf{Coordination, cooperation, and defection scenarios and the role of hindrance to change.} 
 A) Impact of different scenarios of coordination, cooperation and defection on maximum [CO$_2$] and years to achieve net-zero emissions with different degrees of carbon pricing and financial incentives. In the box plot, the red central marks indicate the median, and the bottom and top edges display the 25$^{\rm{th}}$ and 75$^{\rm{th}}$ percentiles, respectively. The whiskers extend to twice the interquartile range. 
  B) Maximum [CO$_2$] (red line) and years to achieve net-zero emissions (blue line) as a function of the socio-political resistance to decarbonization quantified by the block-specific parameters $\mu_i$. A $\mu_i$, randomly extracted from a Gaussian distribution with increasing mean (from  $\mu_i \approx 2.5 - 4.5 \times 10^{-2}$), was assigned to each block. The trends and error bars represent, respectively, the binned mean and standard deviation of possible realizations.  
  Panel (A) was obtained for fixed socio-political hindrance $\mu$ ($\mu_1$=$\mu_2$=$\mu_3$=$\mu_4$), but varying carbon pricing policies ($\tau_i$). On the contrary,  Panel (B) was derived for fixed $\tau$ ($\tau_1$=$\tau_2$=$\tau_3$=$\tau_4$) and variable $\mu_i$.
}
  \label{fig:Fig_3}
\end{figure}
%%%%%%%%%%%%%% End Figure 3 %%%%%%%%%%%%%%%%%%%%%
%%

We investigate these questions by simulating different coordination, cooperation, and defection scenarios within our minimalistic energy balance - CO$_2$ emission model. For simplicity and as a general example, we classify countries into four  blocks based on their propensity to invest in clean energy and projected energy demand (see Table \ref{Tab:Table_1}). For instance, Block 1 encompasses countries with low projections in energy demand increase and high propensity in promoting clean energy (i.e., low hindrance). 
On the other edge of the spectrum, Block 4 includes those countries with rising energy demand and high hindrance to investing in low-carbon energy production. We assume that Blocks 1 and 2 are keen to coordinate a global effort and include green SPS, while conservative SPS in Blocks 3 and 4 are more prone to cooperate, investing less or even defect Blocks 1 and 2. Similar classifications have been proposed in the framework of Climate Clubs, defined as groups of countries that share carbon policies and enforce penalties for those that do not comply \citep{keohane2017toward,nordhaus2020climate}. These may also be building blocks for a polycentric approach for coping with climate change \citep{ostrom2014polycentric}.

Figure \ref{fig:Fig_3}A shows the possible impact of coordination, cooperation, and defection on the time to reach NZ (years to NZ; $t_{\rm{NZ}}$) and peak in CO$_2$ concentration. 
Different degrees of commitment are reproduced by simulating various policies of financial incentives and carbon pricing (see Methods). As expected, rapid achievement of net-zero emissions is only possible in a scenario of deep coordination. There, the peak in CO$_2$ and the time to reach NZ strongly vary as a function of the effort in boosting clean energy investments. A cooperation regime can still lead to emissions cuts and NZ. However, in this case, the margins for a rapid transition to clean energy are limited, even when Blocks 1 and 2 invest heavily. 
Overall, the spectrum of possible pathways to NZ reproduces what Barrett and Dannenberg \citep{barrett2012climate} qualitatively described as scenarios of catastrophe avoidance. Depending on the uncertainty about the occurrence of the catastrophe, either a coordination game (coordination) or a prisoners’ dilemma (cooperation and defection) may emerge. In our model, the uncertainty about the effects of climate change is reflected in the perceived economic damage and the hindrance to change $\mu$. 

These scenarios provide useful insights on how different strategies may impact the path to NZ, assuming that all SPSs involved have the same hindrance to climate mitigation actions. To investigate how the block-specific sensitivity to climate issues could affect the achievement of NZ, we simulate a combination of coordination scenarios in which each block is characterized by a different hindrance to change. The results (Figure \ref{fig:Fig_3}B) highlight that even with deep international collaboration and effective policies, strong socio-political hindrance limits the possibility to reach NZ in short periods and before crossing high concentrations of carbon dioxide. %to shifting from a carbon- to a clean-energy-based economy 
Counterintuitively, homogeneity in climate change risk perception does not necessarily translate into lower emissions. For fixed average hindrance ($\langle \mu \rangle = const$), a coordination scenario where some blocks exert less hindrance (i.e., below average) than others (i.e., above average) may lead to lower concentrations. This behavior can be explained by the different responses of dissimilar SPSs to rising emissions and clean energy investments. If the additional investment the green SPS is willing to pay to cut emissions exceeds the missing investments of the conservative SPS, the total share of clean energy is higher than in the case in which everyone has the same risk perception.
%======================%
%    Conclusions
%=======================%
\section*{Conclusions}
\label{sec:Conclusion}
Climate change is frequently referred to as the greatest tragedy of the commons of our socio-political systems \citep{battersby2017news}. Selfish and short-sighted economic interests are pushing the Earth’s climate towards a warm stage considered irreversible by many \citep{rockstrom2009planetary}. Since the 1990s, most governments have come to the understanding that shared policies and protocols are essential to mitigate climate change, and several agreements have been signed \citep{protocol1997kyoto,agreement2015paris, keohane2016cooperation}. Yet, global collaborative climate policies have had limited application so far, as they face the challenge of contrasting priorities of individuals and local entities. 

Socio-political acceptance has a key role in defining the path to net-zero emissions.
If, from a macro-economic and climate change risk perspective, multiple pathways are plausible \citep{seneviratne2018many,riahi2021cost,biermann2012navigating}, the socio-political feedback could be the double-edged sword determining whether mitigation actions are self-reinforcing or self-undermining.    
Our findings demonstrate that the perception of climate change risk and local and international policies aimed at boosting clean energy investments are profoundly interconnected and can significantly impact near- and long-term CO$_2$ trajectories. Therefore, it is possible that the socio-political feedbacks on the path to net zero could influence the trajectories of the Earth System for decades to centuries and beyond \citep{steffen2018trajectories}.
%================================================================%
\section*{Data and Methods}
%================================================================%
% \matmethods{\vspace{-15pt}
\subsection*{Model design.}
Our model is designed to parsimoniously capture the interplay of social dynamics, mitigation strategies, and global CO$_2$ concentration ([CO$_2$]) fluctuations (see Figure \ref{fig:Fig_1}). It describes the balance between fossil fuels emissions, natural sinks, and offset technologies as the aggregated contribution of groups of countries or sub-regional entities. Following similar use in the literature, each group represents a block \textit{i}, characterized by shared policies and socio-political acceptance of clean energy investments. Similar classifications -- for instance, the Carbon Clubs \citep{keohane2017toward,nordhaus2020climate} -- have been proposed in the literature to represent groups of governments with the same incentive and penalty structure aimed at discouraging free-riding behaviors. Another example comes from the experience of the so called C40 network, which include nearly 100 major cities worldwide collaborating to take common climate actions and become carbon neutral \citep{acuto2016give}.

Based on empirical evidence from synthesis data \citep{friedlingstein2021global}, we assume that the energy demand ($E_{0}$) is the major driver for global anthropogenic emission. For each block, $E_{0,i}$ is met by a combination of CO$_2$-emitting energy production systems based on fossil fuels ($E_{\rm{FF},\textit{i}}$) and non-emitting technologies, generally referred to as clean energy systems $E_{\rm{CL},\textit{i}}$: $E_{0,i}(t) = E_{\rm{FF},\textit{i}}(t) + E_{\rm{CL},\textit{i}}(t)$.
% \begin{linenomath*}
% \begin{equation}\label{eq:E0}
% E_{0}(t) = E_{FF}(t) + E_{\rm{CL}}(t).
% \end{equation}
% \end{linenomath*}
The CO$_2$ emissions related to energy production are in part balanced by the natural sinks NS$_i$, defined as the flux from the terrestrial and marine ecosystem and incorporating land-use change emissions. Other sinks include various technologies used to offset carbon emissions such as carbon capture and storage  \citep{gibbins2008carbon,boot2014carbon}, land-based climate solutions \citep{griscom2017natural,frank2021land}, and direct atmospheric carbon capture \citep{sanz2016direct,marcucci2017road}. We refer to these mitigation strategies collectively as offset (OF$_i$).

The transition to $E_{\rm{CL},\textit{i}}$, pushed by the rising concern about climate change, is expected to reduce CO$_2$ emissions \citep{bang2010energy,kellogg2019climate}. How much each block invests in clean energy, therefore reducing $E_{\rm{FF},\textit{i}}$, depends on socio-political acceptance, represented, for each block, by the variable SA$_i$ \citep{wustenhagen2007social,sovacool2014we}. As explained in the following sections, our estimate of the perturbation of the global carbon budget explicitly accounts for the SA$_i$ feedback through a parametrization of the perceived impact of climate change and growing $E_{\rm{CL},\textit{i}}$ on clean energy investments $I_{\rm{CL},\textit{i}}$.  
%============================================%
%      Methods: Minimalistic NZ model
%============================================%
\subsection*{Global trajectories of CO$_2$ emissions.}
\label{sec:CO2Model}
The global carbon balance can be written as a function of the cumulative contribution to emissions (or sinks) of each block, 
% \begin{linenomath*}
\begin{equation}\label{eq:Co_Blocks2}
\frac{d\, [{\rm CO}_2]}{dt} = \sum_{i=1}^{n} \{ \underbrace{\eta_i [E_{0,i}(t) - E_{\rm{CL},\textit{i}}(t)]}_\text{Anthr. Emissions}  - \underbrace{\rm{OF}_\textit{i}(\textit{t}) }_\text{Anthr. Sinks} -  \underbrace{ \rm{NS}_\textit{i}(\textit{t}, [{\rm CO}_2])}_\text{Net Natural Sinks} \}, 
\end{equation}
% \end{linenomath*}
where $\eta_i$ represents the increase in atmospheric CO$_2$ concentration per unit of energy generated by fossil fuels, and $E_{0,i}(t) - E_{CL,i}(t)$ is the portion of energy demand that cannot be accommodated by clean energy at the time $t$ and is met by CO$_2$-emitting fossil fuels.

The natural sinks $\rm{{NS}}_\textit{i}(\textit{t}, [{\rm CO}_2])$ typically rise with increasing [CO$_2$], as carbon fertilisation effect is expected to increase the carbon uptake of most terrestrial and marine ecosystems \citep{cao1998dynamic,adejuwon2012rainfall}. This increase in $\rm{NS}_\textit{i}$ with CO$_2$ may significantly decelerate when other abiotic or biotic limiting factors related to water and nutrients availability, soil degradation, or maximum photosynthetic capacity become dominant  \citep{oren2001soil,Perri2019,Perri2020,wang2020recent}. In the near future, total natural sinks may also be limited by land-use change and the progressive loss of primary and secondary habitats \citep{gitz2003amplifying,macreadie2013loss,zhongming2021ar6}.
To describe the complex relation between $\rm{NS}_\textit{i}$ and [CO$_2$] in a flexible way, we adopt a power-law function of the form: $\rm{NS}_\textit{i}([\rm{CO}_2])$= $a_i [\rm{CO}_2]$ ${^{b_i}}$. This relation can describe different degrees of non-linearity, with ${a_i}$ and ${b_i}$ capturing the sensitivity of $\rm{NS}_\textit{i}$ to [CO$_2$] and allowing the possibility of its temporal variation. In particular, a value of the coefficient ${b_i}$<1 can account for the decelerating sink rate with [CO$_2$], and ${a_i}$<0 can simulate the shift of natural systems from net sink to net source.

Following recent projections \citep{capuano2018international,dudley2019bp}, the global energy demand in Equation (\ref{eq:Co_Blocks2}) is assumed to increase of nearly 50$\%$ by 2050. The trajectories of global [CO$_2$] will depend on how much of this increasing demand is met by fossil fuels and clean energy sources. 
%%%%%%%%%%%%%%%%%%%%%%%%%%%%%%%%%%%%%%%%%%%%%%%%%%%
%%%%%% Clean energy production and investments %%%%
%%%%%%%%%%%%%%%%%%%%%%%%%%%%%%%%%%%%%%%%%%%%%%%%%%%
\subsection*{Clean energy production and investments.}
\label{sec:ECLModel}
The dynamics of the block-specific clean energy production $E_{\rm{CL}}$ depend on the investment in clean energy $I_{\rm{CL},\textit{i}}$ and a depreciation term
% \begin{linenomath*}
\begin{equation}\label{eq:ECL}
\frac{d\, E_{\rm{CL},\textit{i}}}{dt} = \alpha_{i} I_{\rm{CL},\textit{i}}(t) - \beta_{i} E_{\rm{CL},\textit{i}}(t), 
\end{equation}
% \end{linenomath*}
with $\alpha_{i}$ being the energy produced per unit of investment, $\beta_{i}$ a depreciation coefficient accounting for efficiency loss of old plants and maintenance costs.

$I_{\rm{CL},\textit{i}}$ is highly variable across blocks, and it depends on carbon pricing policies and how acceptable it is for a socio-economical system to invest in clean energy \citep{moula2013researching,nathwani2019affordable,jan2020social,constantino2021source}.
The temporal variation of $I_{\rm{CL},\textit{i}}$ may be written as
% \begin{linenomath*}
\begin{equation}\label{eq:ICL}
\frac{d\, I_{\rm{CL},\textit{i}}}{dt} = \tau_i \rm{SA}_i(\textit{t}, [{\rm CO}_2]), 
\end{equation}
% \end{linenomath*}
where $\tau_i$ is a parameter quantifying the propensity of stakeholders to invest in clean energy for a given level of dimensionless socio-political acceptability SA$_i (\textit{t}, [{\rm CO}_2])$.
Such acceptability can change over time because it is predicted to increase with rising [CO$_2$] and decrease with growing $E_{\rm{CL},\textit{i}}$. The economic damage caused by global warming and extreme events attributed to climate change can shift the public opinion and consumption patterns to more sustainable technologies \citep{nisbet2007polls,nath2011critical}. 
At the same time, SA$_i (\textit{t}, [{\rm CO}_2])$ is expected to be negatively correlated with $E_{\rm{CL},\textit{i}}$ because of the fossil fuels-related job loss associated with the clean energy transition and the increasing perceived impact of large-scale wind and solar energy projects \citep{lesser2010renewable,lambert2012challenges}. Moreover, once clean energy takes over the most technically and economically viable portion of the demand, further emission cuts from sectors such as aviation and many industrial applications becomes challenging and less acceptable \citep{davis2018net,papadis2020challenges}.

This intrinsic resistance to decarbonization is here encapsulated in the hindrance parameter $\mu_i$. 
On the one hand, $\mu_i$ is expected to decrease with the economic damage of rising [CO$_2$]; on the other hand, $\mu_i$ should increase in fossil fuel-dependent SPS, where clean energy investments are associated with job loss and negative impacts on local economies.  Social acceptability is thus estimated as
% \begin{linenomath*}
\begin{equation}\label{eq:SA}
\rm{SA}_\textit{i}= \left[ \rm{PED}_\textit{i}(\textit{t}, [\rm{CO}_2]) - \mu_\textit{i} \textit{E}_{\rm{CL},\textit{i}} (\textit{t}) \right] ^{\textit{d}_\textit{i}},
\end{equation}
% \end{linenomath*}
where $\textit{d}_\textit{i}$ can account for non-linearities in the socio-political response to damage and clean energy, $\rm{PED}_\textit{i}([\rm{CO}_2])$ is the perceived economic damage of climate change that can be approximated by a quadratic function of temperature change \citep{nordhaus2017revisiting}, which, in turn, grows with $[\rm{CO}_2]$ following a logarithmic law \citep{houghton1997introduction,joos2001global}.
By substituting this logarithmic law into the damage function of Nordhaus' DICE model \citep{nordhaus1993optimal,nordhaus2017revisiting}, the perceived economic damages of climate change can be described through
% \begin{linenomath*}
\begin{equation}\label{eq:Damage}
\rm{PED}_\textit{i}([\rm{CO}_2]) = \phi_{1,\textit{i}} ln\left(
\frac{[\rm{CO}_2]}{[\rm{CO}_2]_{\rm{Ref}}}\right)
+ \phi_{2,\textit{i}} ln^2\left(
\frac{[\rm{CO}_2]}{[\rm{CO}_2]_{\rm{Ref}}}\right),
\end{equation}
% \end{linenomath*}
where $\phi_1$ and $\phi_2$ are coefficients that quantify the relative economic impact of changes in the radiative forcing due to changes in $[\rm{CO}_2]$ with respect to the reference value. In spite of its apparent complexity, Equation (\ref{eq:Damage}) describes a nearly linear relation between damage and relative $[\rm{CO}_2]$ increments. It is also important to underline that the economic damage is highly variable across regions due to different vulnerabilities to climate change \citep{field2014climate}. 
Although all block-level emissions contribute to rising global $[\rm{CO}_2]$, the inequality of climate change complicates international cooperation and may be a further incentive for free-riding \citep{mahlstein2011early,king2018inequality}. For example, if a fossil-fuel-based block does not suffer substantial economic damage due to its carbon emissions, it is less likely to invest in clean energy (low SA$_i$). This may lead to increasing $[\rm{CO}_2]$, for which a more vulnerable block would pay the price in terms of its economical damage or investing in clean energy. 
% Population and ecosystems in the tropics are known to be the most vulnerable to climate change, especially island nation exposed to sea-level rise. 

The system of Equations (\ref{eq:Co_Blocks2})-(\ref{eq:Damage}) thus captures various aspects of the complex natural-human system that define the feedback between socio-political acceptance of decarbonization and carbon emissions. The model can describe how each block responds to global $[\rm{CO}_2]$ and local energy demand as a function of carbon pricing policies, expected economic damage due to climate change, and hindrance to change. Moreover, it accounts for the strategic interaction between blocks. The degree to which each block invests to achieve the common NZ goal also depends on other blocks' actions through global carbon emissions.
%%%%%%%%%%%%%%%%%%%%%%%%%%%%%%%%%%%%%%%%%%%%%%%%%%%
%%%%%% Clean energy production and investments %%%%
%%%%%%%%%%%%%%%%%%%%%%%%%%%%%%%%%%%%%%%%%%%%%%%%%%%
\subsection*{Equilibrium and Stability analysis.}
\label{sec:Steady-state}
Equations (\ref{eq:Co_Blocks2}-\ref{eq:ICL}) represent a 3-dimensional dynamical system with a unique interior equilibrium point. 
Linearizing total sinks about an average representative value ($\{ \rm{NS}_\textit{i}(\textit{t}, [{\rm CO}_2]) + \rm{OF} \}_{\rm{Linear}}= \gamma_1 + \gamma_2[{\rm CO}_2]$), we can derive a simple expression of this equilibrium point in case of $n$=1 (deep coordination)
% \begin{linenomath*} 
\begin{equation}
\label{EqPoint}
 \left\{
    \begin{array}{ll}
[{\rm CO}_2]_{\rm{eq}}=\frac{\left(E_0-\gamma _1\right) \mu }{\gamma _2 \mu +\eta }\\
E_{\rm CL, eq}=\frac{E_0-\gamma _1}{\gamma _2 \mu +\eta }\\
I_{\rm CL, eq}=\frac{\beta  \left(E_0-\gamma _1\right)}{\alpha  \left(\gamma _2 \mu +\eta \right)}\\
 \end{array}
  \right..
\end{equation}
% \end{linenomath*}
A linear dependence between NS and [CO$_2$] remains realistic for narrow concentration intervals. 

We performed a linear stability analysis of this equilibrium point \citep{strogatz2015nonlinear}. The system behaves as a damped oscillator with three degrees of freedom ([CO$_2$], $E_{\rm{CL}}$, and $I_{\rm{CL}}$) and predisposition to overshooting \citep{bhatia2002stability}. The eigenvalues are, in fact, $\lambda_1  \in \ \mathbb{R}$, $\lambda_2 = \lambda_3  \in \mathbb{C}$, all with negative real part. 
The system has one stable equilibrium point ($[{\rm CO}_2]_{\rm{eq}}$, $E_{\rm{CL,eq}}$, and $I_{\rm{CL,eq}}$), meaning that it tends to return to this point after any disturbance. In particular, after a perturbation, due, for instance, to an increase in [CO$_2$] resulting from rising anthropogenic emissions, the system tends to oscillate around its equilibrium with decreasing amplitude (underdamped oscillator).

Moreover, as an example of the possible consequence of diverging hindrances to change, we can solve (\ref{eq:Co_Blocks2}-\ref{eq:ICL}) in case of two blocks ($n$=2) for which the steady-state yields $[{\rm CO}_2]_{\rm{eq}} = \frac{(E_0 - \gamma_1) \mu _1 \mu _2}{\gamma_2 \mu _1 \mu _2+\eta  \left(\mu _1+\mu _2\right)}$. An interesting characteristic of this stable point is that maintaining a constant average resistance ($\langle \mu \rangle$=$(\mu _1 + \mu _2)/2$= Const) but varying $\mu _1$ and $\mu _2$ accordingly results in a $[{\rm CO}_2]_{\rm{eq}}$ lower than on the case of $\mu _1$=$\mu _2$. The $[{\rm CO}_2]_{\rm{eq}}$  decreases as the difference between the two resistance parameters increases. The implication of this finding is that, contrary to our common perception, a homogeneous perception of climate change risk ($\mu _1$ = $\mu _2$) may not lead to higher investments and lower emissions in the long run. Divergent perceptions (e.g., $\mu _1 > \mu _2$) result, instead, in lower stable concentration, with $[{\rm CO}_2]_{\rm{eq}}$ rapidly decreasing as the difference between $\mu _1$ and $\mu _2$ grows (see Extended Data Figure 2). However, this scenario is possible only assuming that block 1 can compensate for the missing effort of block 2, and it does not necessarily represent the best trajectory to NZ. Although divergent hindrance to change may result in a lower carbon dioxide concentration at equilibrium, it would also imply a longer time to achieve NZ and a higher peak in $[{\rm CO}_2]$ (see Figure \ref{fig:Fig_3}).   
%=======================================================%
%      Methods: Dishomogeneous effort to achieve NZ
%=======================================================%
\subsection*{Scenarios of coordination, cooperation, and defection.}
\label{sec:Blocks}
Although nearly all countries have signed the Paris Agreement that sets shared goals for a clean energy transition \citep{agreement2015paris}, most of them are failing to meet such ambitious targets. 
The transition to clean energy is particularly problematic for developing countries, where the increasing energy demand and limited resources call for cheap and readily available energy sources.
Given the different projected increment in energy demand \citep{dudley2019bp} and propensity to invest in clean energy \citep{kunreuther2014integrated,ang2020individualism}, we grouped countries worldwide into four ideal groups or blocks (Table \ref{Tab:Table_1}): Block 1: developed countries transitioning to clean energy;  Block 2: developing countries investing in clean energy;  Block 3: developed countries relying on fossil fuels;  Block 4: developing countries relying on fossil fuels. % (See Figure \ref{fig:Fig_4}a). 

The scenarios of coordination, cooperation and defection in Figure \ref{fig:Fig_3}A were obtained as follows. We consider, as a general example, four blocks $i$ with the same initial energy demand, natural sinks, fossil fuels conversion factor, and share of clean energy, but different projected $E_{0,i}(t)$ and commitment to invest in clean energy (i.e., different $\tau_i$). 
Blocks 1 and 3 are assumed to maintain a constant energy demand, while blocks 2 and 4 experience a doubling of the initial $E_{0,i}(t)$ following a non-linear saturation trend. A high $\tau_i$ is assumed for blocks 1 and 2, whereas blocks 3 and 4 rely primarily on fossil fuels and possess a low $\tau_i$. As displayed in Table \ref{Tab:Table_1}, blocks 1 and 3 could represent developed countries for which energy demand is projected to remain constant or only marginally increase in the next decades, while blocks 2 and 4 encompass developing countries with fast-growing energy consumption \citep{dudley2019bp}. Using equations (\ref{eq:Co_Blocks2})-(\ref{eq:ICL}) we describe the dynamics of global atmospheric [CO$_2$] in response to the different investment in clean energy of the four blocks.

The role of the hindrance to change, described here through the sensitivity of $\rm{SA}_i$ to increasing $E_{\rm{CL}_i}$, was inferred by simulating the global carbon dioxide trajectories as a function of $\mu$, obtained extracting a random combination of four $\mu_i$ (see  Figure \ref{fig:Fig_3}B). The values of $\mu_i$ were chosen to fluctuate around increasing mean values and within the range $2.5 \times 10^{-2} \leq \mu_i  \leq 4.5 \times 10^{-2}$.
%======================%
%        Tables
%======================%
%%%%%%%%%%%%%%%%%%%%%%%%%%%%%% Table 1
\begin{table}
\caption{Blocks definition based on their projected energy demand and propensity for climate change mitigation policies.}
  \begin{center}
\begin{tabular}{  c || c | c  }
 \hline
\textbf{ } & \thead{Low Increment in \\ Energy Demand}   & \thead{High Increment in \\ Energy Demand}\\
 \hline 
\thead{High \\ Propensity \\ for Clean \\ Energy}  &  \makecell{\textbf{Block 1} \\ Developed countries \\ transitioning to $E_{\rm{CL}}$\\  (low $\mu$, low $E_0$, high $\tau$)} & \makecell{\textbf{Block 2 }\\ Developing countries \\ transitioning to $E_{\rm{CL}}$\\ (low $\mu$, high $E_0$, high $\tau$)} \\  \hline
\thead{Low \\ Propensity \\ for Clean \\ Energy}  &  \makecell{\textbf{Block 3} \\ Developed countries \\ relying on fossil fuels\\ (high $\mu$, low $E_0$, low $\tau$)}  & \makecell{\textbf{Block 4} \\ Developing countries \\ relying on fossil fuels\\ (high $\mu$, high $E_0$, low $\tau$)} \\
 \hline
\end{tabular} 
  \end{center}
 \label{Tab:Table_1}
\end{table} 
\clearpage
% \showmatmethods{} % Display the Materials and Methods section
\subsubsection*{Acknowledgment}{\sf\small Funding in support of this research was provided by Princeton University’s Dean for Research, High Meadows Environmental Institute, Andlinger Center for Energy and the Environment, and the Office of the Provost International Fund. We also acknowledge support from the US National Science Foundation (NSF) grant nos. {EAR-1331846} and EAR-1338694, the BP through the Carbon Mitigation Initiative (CMI) at Princeton University, the Moore Foundation, and the European Research Council Advanced Grant project ERA (ERC-2016-ADG-743080).
}

\end{document}